# Recent progress in the theory of random surfaces and simplicial quantum gravity

J. Ambjørn[a]

[a]The Niels Bohr Institute,
Blegdamsvej 17, DK-2100 Copenhagen Ø, Denmark

Some of the recent developments in the theory of random surfaces and simplicial quantum gravity is reviewed. For 2d quantum gravity this includes the failure of Regge calculus, our improved understanding of the $c > 1$ regime, some surprises for q-state Potts models with $q > 4$, attempts to use renormalization group techniques, new critical behavior of random surface models with extrinsic curvature and improved algorithms. For simplicial quantum gravity in higher dimensions it includes a discussion of the exponential entropy bound needed for the models to be well defined, the question of "computational ergodicity" and the question of how to extract continuum behavior from the lattice simulations.

## 1. INTRODUCTION

The first attempt to discretize gravity in a coordinate free way dates back to the seminal work of Regge from 1961 [1]. The purpose of Regge was purely classical. He wanted to approximate a manifold with a given metric by a piecewise linear manifold. Regge made the important observation that parallel transport and the integral of curvature have a natural definition on piecewise linear manifolds. For a given abstract triangulation an assignment of length to the links, consistent with the triangle inequalities, will define uniquely the concept of curvature and parallel transport in a way which coincides with an embedding of the triangulation as a piecewise linear manifold in some ambient space $R^n$. The assignment of integrated curvature is as follows: Let $T$ be the triangulation, let $\{n_i\}$ denote the set of $i$-simplexes in $T$. A given $(d-2)$-simplex $n_{d-2}$ will belong to a number $o(n_{d-2})$ of $d$-simplexes, $n_d(k)$, $k = 1, 2, ..., o(n_{d-2})$, where $o(n_{d-2})$ is called *the order* of $n_{d-2}$. For one of the $n_d(k)$'s we have that $n_{d-2}$ belongs to precisely two $(d-1)$-subsimplexes and if we denote the angle between these two subsimplexes $\theta_k$, the *deficit angle* $\varepsilon(n_{d-2})$ is defined as

$$\varepsilon(n_{d-2}) = 2\pi - \sum_{k=1}^{o(n_{d-2})} \theta_k. \tag{1}$$

The angle $\theta_k$ is defined in the ambient space $R^n$ but can be expressed entirely in terms of the length of the links of the abstract triangulation. In the same way the volume of $V_i(n_i)$ of an $i$-simplex in the triangulation can be expressed entirely in terms of the length of the links. The fundamental definitions of Regge, which is in accordance with the concept of parallel transport on the piecewise linear manifold is given by:

$$\int_M d^d\xi \, \sqrt{g} \to \sum_{n_d} V_d(n_d) \tag{2}$$

$$\int_M d^d\xi \, \sqrt{g}\, R = \sum_{n_{d-2}} \varepsilon(n_{d-2})\, V_{d-2}(n_{d-2}). \tag{3}$$

The idea of Regge was to consider the link length as the classical dynamical variables. For a given manifold one should choose a suitable triangulation and by varying the length of the links find the extremum of the Einstein action. The corresponding piecewise linear manifold should then qualify as a approximation to the underlying smooth manifold which is a solution the Einstein equations. This makes perfect sense at a classical level since the lengths of the links are invariants and for a sufficiently fine triangulation one expects a good approximation to the smooth manifold.

Inspired by lattice gauge theories the quantum analogue to Regge calculus was suggested. The simplest idea is to use the classical link variables and integrate over these variables. However, one encounter a complicated counting problem. In quantum gravity we want to integrate over equivalence classes of metrics. For a given abstract triangulation there can be many length assignments which correspond the same equivalence class of metrics as is clearly seen by considering triangulations of the plane with the usual Euclidean metric. For a given triangulation of the plane there is usually plenty of room for moving the vertices around thereby changing the length of the links without changing the Euclidean metric of the plane. We conclude that the replacement

$$\int_{\mathcal{M}} \frac{\mathcal{D}g_{ab}}{\text{Vol}(\text{diff})} \to \int \prod_i dl_i J(l_i) \quad (4)$$

involves as highly non-trivial jacobian $J(l_i)$. In eq. (4) the functional integration on the lhs is over equivalence classes of triangulations, symbolized by the division by the "volume" of the diffeomorphism group. The integration on the rhs of eq. (4) is over all link length compatible with the triangle inequalities. We can view the triangle inequality as incorporated in the Jacobian $J(l_i)$. An important aspect of the "quantum version" of the Regge calculus is the absence of a cut-off. As formulated above one has discretized gravity but not introduced a cut-off. In general one will have to introduce a smallest distance or higher derivative terms in the discretized action in order to make the integration over link length well defined.

A different approach was advocated by Weingarten, both for strings and for gravity. The idea was to consider manifolds embedded in hypercubic lattices. In this case we have a natural cut off, the lattice spacing $a$, and we have no problem with overcounting since each manifold is explicitly given on lattice as a "physical", coordinate independent manifold. The metric assignment for such a piecewise linear surface can be made by the Regge calculus. Since all building blocks are identical it will only depend on the number of $d$- and $(d-2)$-dimensional building blocks of the $d$ dimensional manifold on the $D$-dimensional hypercubic lattice. In this way we get:

$$S_{eh}[G, \lambda] = \int d^d\xi \sqrt{g} - \frac{1}{16\pi} \int d^d\xi \sqrt{g} R \to$$
$$S[k_{d-2}, k_d] = k_d N_d - k_{d-2} N_{d-2} \quad (5)$$

and the partition function will be given by

$$Z[k_{d-2}, k_d] = \sum_{M \in Z^D} e^{-S[k_{d-2}, k_d]}. \quad (6)$$

From this point of view the hypercubic manifolds form a grid in the space of equivalence classes of metrics. The question is to which extent it is uniform and dense in the limit of infinite $N_d$. While a dependence on the dimension $D$ of the hypercubic lattice is natural in the case of string theories, it is not desirable in the case of gravity. Weingarten suggested a formulation independent of the underlying hypercubic lattice, but the more streamlined formulation in terms of abstract so-called dynamical triangulations were first given in two dimensions in [2–4] and in higher dimensions in [13,6,7]. In this formulation the summation over hypercubic manifolds is replaced by the summation over abstract $d$-dimensional triangulations where the length of the links is taken to be $a$. In this way we still have a cut off which is "reparametrization invariant" and by Regge calculus we can still assign a metric and the action (5) to the given abstract triangulation. Again the action is only a function of $N_{d-2}$ and $N_d$ since all $k$-simplexes in the triangulation are identical. Each triangulation is in this way a representative of a whole equivalence class of metrics. There will be no problem with overcounting and by summing over all abstract triangulations we again lay out a grid in the space of equivalence classes of metrics. Eq. (6) will be replaced by

$$Z[k_{d-2}, k_d] = \sum_{T \in \mathcal{T}} e^{-k_d N_d + k_{d-2} N_{d-2}} \quad (7)$$

Eq. (7) describes a discretization of pure gravity. $T$ denotes an abstract triangulation in a given class of triangulations $\mathcal{T}$. In addition one can couple matter to gravity in a natural way [8]. Since the link length is taken to be one it is easy to show that a Gaussian scalar field couples as follows:

$$S[\phi, g] = \int_M d^d\xi \sqrt{g}\, g^{ab}\partial_a\phi\partial_b\phi \to$$

$$S[\phi, T] \sum_{(ij) \in T} (\phi_i - \phi_j)^2 \qquad (8)$$

In this formula $i$ and $j$ denote vertices and $(ij)$ links in the triangulation $T$. This formula can be derived by assuming that the field in the interior is given by the linear extrapolation of the field value $\phi_i$ at the vertices. Space-time in the interior of the $d$-simplexes is assumed flat and is dictated by the length assignment $a$ to all links. Alternatively one can place the discretized field values in the center of the $d$-simplexes and again arrive at an expression like (8), only is the index $i$ now a $d$-simplex and the summation is over neighboring $d$-simplexes. Spin systems and even gauge fields can also be coupled to gravity in this formalism.

It is possible to divide the study of simplicial quantum gravity in two classes: the study of two-dimensional quantum gravity and higher dimensional gravity. The theory of two-dimensional gravity is well defined if we restrict the topology of the manifold. It also includes the theory of bosonic strings in $R^D$ since such strings can be viewed as two-dimensional gravity coupled to $D$ scalar fields. It is also possible to move in the direction of the theory of membranes since one can introduce extrinsic curvature terms in such models. For the two-dimensional theories the situation is very nice from the point of view of computer simulations: We have systems where all our methods: Monte Carlo simulations, finite size scaling etc. seem to work well. This is illustrated in fig. 1 and fig. 2. The results presented are from a Monte Carlo simulation of the Ising model coupled to two-dimensional gravity. The simplest finite size scaling tells us that

$$M_N \sim N^{-\beta/\nu d_H}, \qquad \chi_N \sim N^{\gamma/\nu d_H} \qquad (9)$$

where $N$ is the volume (i.e. in this case the number of triangles in the triangulation), $M_N$ the magnetization and $\chi_N$ the susceptibility at the pseudo critical point corresponding to $N$. $d_H$ denotes the Hausdorff dimension of two-dimensional gravity coupled to a $c = 1/2$ conformal field theory. From eq. (9) we get

$$\frac{\beta}{\nu d_H}\big|_N \equiv -\frac{\log M_N}{\log N}, \qquad \frac{\gamma}{\nu d_H}\big|_N \equiv \frac{\log \chi_N}{\log N}. \qquad (10)$$

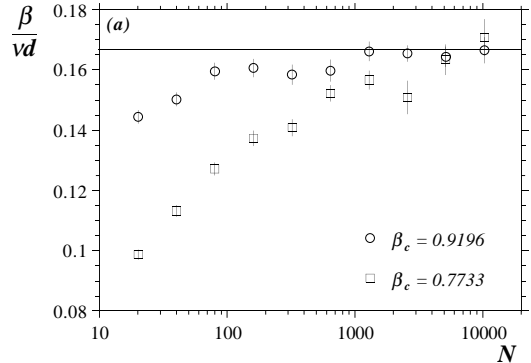

Figure 1. The magnetic exponent as a function of volume as defined by eq. (10).

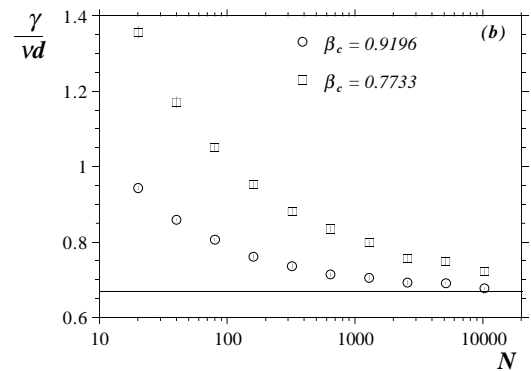

Figure 2. The magnetic susceptibility exponent as a function of volume as defined by eq. (10).

A nice convergence to the known theoretical value is seen. For details I refer to [9].

In addition there are regions, most notably matter fields with central charge $c > 1$ where the theory is not yet understood from an analytical point of view. The numerical results might serve as an important inspiration for analytical approaches which so far have been successful for $c < 1$. If we discuss the higher dimensional gravity theories we move into unchartered territory and the study of these theories is at present at the explorative stage.

In the rest of this review I will concentrate on the developments during the last year. For the two-dimensional theories this includes

1. The failure of two-dimensional quantum Regge gravity.

2. Understanding large c.

3. Surprises for large q-state Potts models.

4. Attempts to formulate renormalization group methods.

5. New algorithms

For three- and four-dimensional gravity it includes

1. The exponential bound on the entropy.

2. The question of "computational ergodicity".

3. How to extract physics from the Monte Carlo simulations.

## 2. TWO-DIMENSIONAL GRAVITY AND RANDOM SURFACES

### 2.1. The failure of quantum Regge gravity

While Regge's approach to gravity makes perfect sense in the classical context in which it was invented, it has already been mentioned that use of the formalism in a quantum context involves a non-trivial Jacobian $J(l_i)$ (see eq. (4)) which is presently unknown. Until now it has been assumed, appealing to universality, that the choice of this measure is not of uttermost importance. A scale invariant measure

$$d\mu(l_i) = \prod_i \frac{dl_i}{l_i} \qquad (11)$$

has been used in most simulations. During the last year two large scale simulations have tested whether Regge gravity in $2d$ agrees with continuum calculations and simplicial quantum gravity. *In both cases the result is so far negative..*

The first work to be discussed is by Holm and Janke ([10] and the contribution to Lattice94). They have looked at an Ising model coupled to Regge gravity. They used toroidal topology and a regular lattice with dynamical length assignment of the links. The spin was located at the vertices. The measure was given by eq. (11) and the lattice sizes in the Monte Carlo simulations ranged from $6^2$ to $512^2$. In addition to the Einstein-Hilbert action a term

$$a \int \sqrt{g} R^2 \qquad (12)$$

was added to the action for $a = 0$, 0.1 and 0.001. The result of a very careful finite size scaling analysis of the critical exponents of the Ising model is summarized in table 1. $\alpha$ is the exponent for the specific heat, $\beta$ and $\gamma$ was defined above, while $\delta$ is the exponent defined by $M \sim H^\delta$ at the critical point. Finally $\nu$ is the exponent which governs the divergence of the correlation length as one approaches the critical point. It is clear from table 1 that the agreement with the standard Onsager exponents for the Ising model on a regular lattice is perfect, while there is disagreement with the known exponents for dynamical triangulated surfaces. Since these latter exponents agrees with Liouville theory one is let to conclude that Regge gravity coupled to Ising models does not describe $c = 1/2$-conformal matter coupled to gravity. It is an interesting question whether it is a fundamental flaw in the approach or it can be repaired by a different choice of measure (11).

The second attempt to test the viability of Regge gravity in the quantum regime is by Bock and Vink (see [11] and contribution to Lattice94). They have performed extensive Monte Carlo simulations of the two-dimensional Regge version of quantum gravity. They have included in addition to the Einstein-Hilbert term a $R^2$ as in (12). In this way it is possible by careful scaling arguments to derive a relation between the coupling constant $a$ in (12) and $\gamma_s$, the string susceptibility exponent. One definition of $\gamma_s$ is as an entropy exponent in the partition function. The partition function for a given area $A$ is defined by

$$\begin{aligned} Z(A) &= \int \frac{\mathcal{D}g_{ab}}{\mathrm{Vol(diff)}} e^{-S[g]} \delta(\int d^2\xi \sqrt{g} - A) \\ &\sim A^{\gamma_s - 3} e^{\mu_c A}. \end{aligned} \qquad (13)$$

In this formula $\mu_c$ is a non-universal cut-off while the subleading correction $A^{\gamma_s - 3}$ defines the universal critical exponent. In pure gravity it can by shown, either by the use of Liouville theory or by

Table 1
Critical exponents in the Regge-Ising model

|         | $\alpha$ | $\beta$ | $\gamma$ | $\delta$ | $\delta$ |
|---------|----------|---------|----------|----------|----------|
| DTS     | $-1$     | 0.5     | 2.0      | 5        | -        |
| Onsager | 0        | 0.125   | 1.75     | 15       | 1        |
| Regge   | 0        | 1.26(2) | 1.75(2)  | 14.9(3)  | 1.01(1)  |

the DTS-model, i.e. simplicial quantum gravity, that

$$\gamma_s = 2 - \frac{5}{4}\chi, \tag{14}$$

where $\chi$ is the Euler characteristic of the surface. While $\gamma_s$ is readily identified in simplicial quantum gravity one has to use scaling arguments involving the $R^2$ term in the Regge formulation. The problem with the measure is here highlighted as it turns out that $\gamma_s$ seems to depend on choice of measure: If (11) is modified to

$$d\mu_\eta(l) = \prod_i \frac{dl_i^2}{l_i^{2+\eta}}, \tag{15}$$

there will be a dependence on $\eta$. The interpretation of this is not clear. From universality and the principle of ultra-locality one would not expect a change like (15) to affect any continuum quantity. To be conservative Bock and Vink did their simulations with $\eta = 0$ in order to be in agreement with earlier measurements of Hamber where agreement with (14) have been reported. However the improved simulations of Bock and Vink for the sphere and the bi-torus ($\chi = 2$ and $-2$) showed no relation between the Regge theory and the correct results given by eq. (14). Again the conclusion is that we know presently no measure $d\mu(l)$ of Regge gravity which can produce the correct continuum results.

### 2.2. Matter with c>1 coupled to 2d gravity

In the context of statistical mechanics it is easy to couple matter with a central charge $c > 1$ to two-dimensional gravity: Take the statistical system, e.g. a multiple Ising system. It is defined on a regular lattice where it has a critical point and a second order transition and an associated divergent correlation length. The central charge and the associated critical exponents are defined at the critical point. The operational steps needed in order to couple the system to 2d quantum gravity is as follows: Define the statistical system on an arbitrary random lattice which represents a triangulation (or more generally a polygon net) on a surface of fixed topology[1]. Define a new statistical system by taking the *annealed* average over the selected class of random triangulations of the surface. The annealed system will still have a critical point. The order of the transition can change (if it change it is usually to a higher order transition, which is intuitively reasonable) and at the new critical point there will be new critical exponents. From the continuum point of view we attribute this change to the interaction with gravity. However, also purely "geometrical" exponents like $\gamma_s$ can change when we perform the summation over matter configurations and afterwards the summation over triangulations with the effective action in part coming from the usual gravity action and in part coming from the integration or summation over matter fields. We view this as the *back reaction* of the quantized matter on gravity.

For $c < 1$ many of the annealed lattice models have been solved explicitly and, as already mentioned above, the critical exponents agree with the continuum calculations using Liouville theory. However, the continuum approach breaks down for $c > 1$. This is illustrated by the so-called KPZ formula for $\gamma_s$:

$$\gamma_s(\chi = 2) = \frac{c - 1 - \sqrt{(c-1)(c-25)}}{12}. \tag{16}$$

It should be noticed that the formula only makes sense for $c \leq 1$ and that $\gamma_s(c) \leq 0$ for these values of $c$. It has been a longstanding question what happens for $c > 1$. While it is not clear how to proceed in the continuum approach, the theory is clearly well defined and unambiguously defined in the statistical mechanics approach. During the last year there has been an increased understanding of the interaction between gravity and matter with $c > 1$. In particular the limit $c \to \infty$ is now well understood by mean field calculations [12–14].

---

[1]Of course the definition should be *reasonable*, i.e. the part of the action which involves nearest neighbor interaction should be in accordance with the *geometry* represented by the random lattice.

Let me briefly describe the simplest derivation of the $c \to \infty$ results: First I appeal to the Monte Carlo results which has been obtained over the last couple of years: They showed that for the annealed average of many Ising models over random triangulations the following is true:

(1): There is a critical (inverse) temperature $\beta_c$ and a phase transition between a magnetized phase (for large $\beta$) and a phase where the total magnetization is zero (for small $\beta$).

(2): For $\beta > \beta_c$ the exponent $\gamma_s$ jumps to $-1/2$, the value for pure gravity. For increasing $c$, i.e. increasing number of Ising models coupled to gravity, the system is increasing fast completely magnetized just above $\beta_c$ which increases logarithmically with $c$.

(3): At $\beta_c$ one has $\gamma_s(\beta_c) > 0$ for sufficiently large $c$.

(4): In a region below $\beta_c$ there seems to be a very strong interaction between that matter system and the spin system. The random surfaces seem strongly deformed compared to typical surfaces present in computer simulations without many spin systems. Again the effects are enhanced with increasing $c$.

(5): For $\beta \to 0$ the exponent $\gamma_s$ is equal $-1/2$, the value of pure gravity.

Points (1) and (2) motivate the mean field approximation for large $c$: Since magnetization is the same all the way from $\beta = \infty$ to $\beta_c$ the spin excitations will just be a selfconsistent iteration of the minimal energy excitation, i.e. just the first term in the usual weak coupling expansion. The major difference between the situation on a dynamical lattice and a fixed regular lattice is that the minimal energy spin excitation on a regular lattice involves the flip of a single spin since this creates the smallest boundary. Not so on a dynamical lattice. Arbitrary large spin clusters can be separated by small boundaries. This is shown in fig. 3 and after having realized this it is also possible to understand intuitively the interaction between spin and geometry: *The spin excitations which will dominate for large $\beta$ and large $c$ will be the ones which deform the geometry into baby universes of alternating spins.* If we consider the one-loop function where the loop just consists of two links we have the selfconsistent equation shown in fig. 3:

$$\begin{aligned} G(\mu, \beta) &= \sum_T e^{-\mu N_T} (1 + e^{-2\beta} G(\mu, \beta))^{L_T} \\ &= \sum_T e^{-\bar{\mu} N_T} = G_0(\bar{\mu}), \end{aligned} \quad (17)$$

where $\bar{\mu} = \mu - \frac{3}{2}\log(1 + e^{-2\beta}G(\mu, \beta))$ since we have that the number of links $L_T = \frac{3}{2} N_T$ for a triangulation. This implies that we can solve for the one-loop function $G(\mu, \beta)$ in terms of the a know function: $G_0(\bar{\mu})$:

$$G(\mu, \beta) = G_0(\bar{\mu}) \quad (18)$$

$$\mu = \bar{\mu} + \frac{3}{2}\log(1 + e^{-2\beta}G_0(\bar{\mu})). \qquad (19)$$

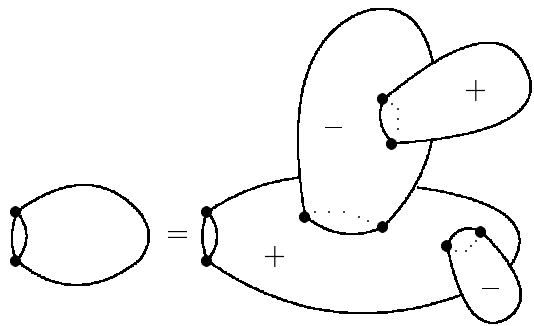

Figure 3. Graphical representation of the mean field equation (17).

The eqs. (18) and (19) are easily solved [13]. The solution for $\gamma_s(\beta)$ is shown in fig. 4.

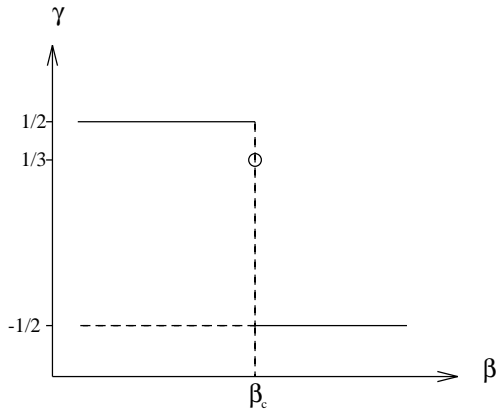

Figure 4. The solution to eqs. (18) and (19) for $\gamma_s$.

The conclusion is that we indeed can obtain $\gamma_s > 0$. In fact one gets $\gamma_s = 1/3$ at the transition point $\beta_c$ and $\gamma_s = 1/2$ below $\beta_s$. It cannot be correct for $\beta \to 0$ according to point 4) above, but it is seen that the mean field approximation explains all other observations. In particular it is tempting to associate the region just below $\beta_c$ with a phase where the interaction between matter and geometry is so strong that the typical surface consists of a number of small baby universes which are individually magnetized but where the total magnetization is zero, i.e. a branched polymer phase. The value $\gamma_s = 1/2$ is precisely the generic value for branched polymers and a surface build of many small baby universes would be such a branched polymer surface. For small $\beta$ the approximation breaks down since there can be many other spin excitations than the selfconstistent iterations of the minimal excitation discussed here.

Is it an accident that the first value of $\gamma_s > 0$ which is different from the branched polymer value is $1/3$ instead of $1/2$? The answer is no! There exists by now a theorem which states that for multiple spin systems coupled to gravity one *must* have

$$\gamma_s > 0 \Rightarrow \gamma_s = \frac{1}{n+1}, \qquad n > 0, \qquad (20)$$

for a spin system. The proof is an elaboration of the eqs. (18) and (19) [15] which leads to the following representation of any $\gamma_s > 0$ in a system with local spin interactions:

$$\gamma_s = \frac{\bar{\gamma}_s}{\bar{\gamma}_s - 1}, \quad \bar{\gamma}_s \leq 0. \qquad (21)$$

In fact the interpretation of (20) and (21) is that the typical surface with $g_s = 1/n + 1$ will consist of baby universes with $\gamma_s = -1/n$ (i.e. $c < 1$) glued together along small loops. The generic value will be $n = 2$ which corresponds to $c = 0$ in which case the interpretation is that the individual baby universes are completely magnetized, i.e. corresponds to $\gamma_s = -1/2$, and the new critical behavior arises at the transition where the spin of individual baby universes are aligned. It should be mentioned that the mean field equations above can be solved in an external magnetic field and one can verify that the transition is a third order transition.

An interesting and yet unanswered question concerns the transition from the branched polymer phase below $\beta_c$ where $\gamma_s = 1/2$ and back to the phase where $\gamma_s = -1/2$, which is valid for small $\gamma_s$.

Another interesting question is how universal this behavior is and if there are examples of nontrivial $\gamma_s >$. By non-trivial I now mean a $\gamma_s$ different from $1/2$ and $1/3$. In fact it is not too difficult to find a meanfield model which is a generalization the model above and which has $\gamma_s = 1/4$ [16]. It is however very interesting that a completely different class of random surface theories seems to have the same critical behavior. This is random surface theories defined on dynamical lattices with the usuall Gaussian action and in addition with an extrinsic curvature term. These have some interest also membrane physics since they can serve as toy models of biological membranes. It has also been conjectured that they are related to superstrings. It is known that they have a phase transion for $D = 3$, $D$ being the dimension of target space, for a finite value of the extrinsic curvature coupling constant [17]. Below the transition point we have the bosonic string which has a non-scaling string tension [18]. At the critical point it seems that the mass and the string tension scales and the random surface model is a candidate for an interesting continuum string theory [19,20]. This has been extensively covered at the former lattice conferences. Using baby universe counting it is possible to measure $\gamma_s$ at the transition point[21]: it seems as if $g_s = 1/4$, i.e. we have an absolutely non-trivial realization of eq. (20). The results of the measurement of $\gamma_s$ at the critical extrinsic curvature coupling is $\gamma_s = 0.27 \pm 0.02$. This is illustrated in fig. 5. It is interesting that the simulations seem to agree with old simulations of on a hyper-cubic lattice where one can also add and extrinsic curvature term [22].

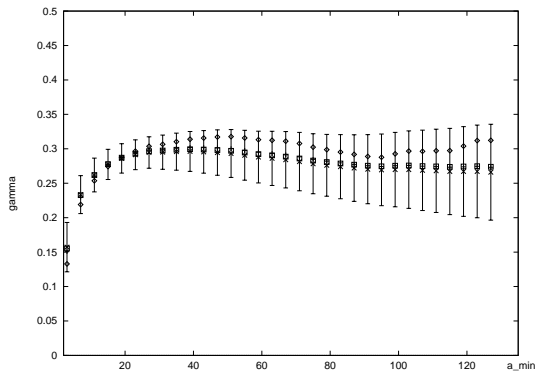

Figure 5. The value of $\gamma_s$ for random surfaces with extrinsic curvature coupling at the critical extrinsic curvature coupling constant. $a_{min}$ denotes the minimal area of the baby universes used in the determination of $\gamma_s$. I refer to [?] for details about the determination of $\gamma_s$ from the the baby universe distributions.

## 2.3. q–state Potts models, q>4

At first it seems that these models should not be interesting since the models on a regular lattice have a *first order* transition, and from this point of view they do not fit into the framework outlined above. However, after coupling to gravity one can prove that large $q \equiv$ large $c$ [12] and in addition it is known that the order of the transition is often decreases when the annealed average over triangulations is performed. For a single Ising model it is known that the transition is decreases from a second order transition on a regular lattice to a third order transition on dynamical lattices. Similarly, since large $q$ corresponds to large $c$ we know from the arguments presented in the last section that $q \to \infty$ also corresponds to a third order transition. What is the situation for finite $q > 4$?

The results of extensive numerical simulations performed for the annealed average of $q = 10$ and $q = 200$ models is shown in the second and third row of table 2 (for details I refer to [9] and the contribution of Thorleifsson at Lattice94). The first row shows the known analytical results for $q = 4$, which corresponds to a $c = 1$ conformal field theory, while the last row shows the mean field results for $c = q = \infty$.

The exponents suggest strongly that $q = 10$ belongs to the same universality class as $q = 4$ while $q = 200$ belongs the same universality class as $q = \infty$. This implies that $\gamma_s = 1/3$ at the transition for $q = 200$ and according the the mean field picture there should be a region below $\beta_c$ where $\gamma_s = 1/2$. The corresponding curve is shown in fig. 6 and gives partial confirmation of this picture despite the fact that it is known to be quite difficult to extract a reliable $\gamma_s$ from the spin models coupled to gravity. An even more convincing confirmation comes from a measurement of the spin clusters. Let

$$W = \frac{b_l + 1}{N_c} \qquad (22)$$

denote the ratio between the number of boundary links $b_l$ between spin clusters and the number $N_c$ of clusters. $W = 1$ only if there is a total magnetization (one cluster) or if there is a tree structure of spin clusters with only one-link boundaries[2]. The measurements of $W$ shows that for $q = 200$ $W$ is close to 1 in a region *below* $\beta_c$. In the same $\beta$ region $N_c$ is large. This shows that in a region below $\beta_c$ there are many small spin clusters. However, the clusters with $b_l = 1$ is in 1-1 correspondence with the underlying geometry of baby universes since we can only have a spin cluster with boundary length one if there is a baby universe separated from the rest of the triangulation by a 1-loop. If we increase $\beta$ and pass $\beta_c$ $W$ will stay very close to one, but $N_c$ will jump from being a fraction of the total number of links to one, corresponding to the situation with total magnetization. For $q = 200$ this is in perfect agreement with the mean field picture. For $q = 10$ there is a clear deviation: for $\beta < \beta_c$ $W > 1$ and it drops to $W = 1$ only after $\beta_c$. This is even more pronounced at $q = 4$ and on regular lattices. $W$ for various Potts model on dynamical triangulated surfaces are shown in fig. 7. To summarize: There seems to be a critical $q_c > 10$ such that for $q < q_c$ the q-state Potts model on dynamical triangulated lattices belongs to the same universality class as $q = 4$, i.e. they describe a $c = 1$ theory coupled to gravity even if the models on a regular lattice cannot be assigned a central charge. For $q > q_c$ the models at the critical point is identical to the meanfield model described in the last section.

Table 2
Critical exponents for the $q$-state Potts models.

|  | $\alpha$ | $\beta$ | $\gamma$ |
|---|---|---|---|
| $q = 4$ | 0 | 1/2 | 1 |
| $q = 10$ | 0.02(4) | 0.53(1) | 1.11(2) |
| $q = 200$ | -0.083(6) | 1.18(4) | 0.31(5) |
| $q = \infty$ | -1 | 1 | – |

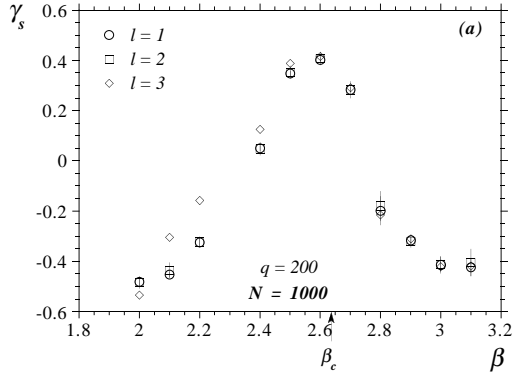

Figure 6. $\gamma_s$ for $q = 200$.

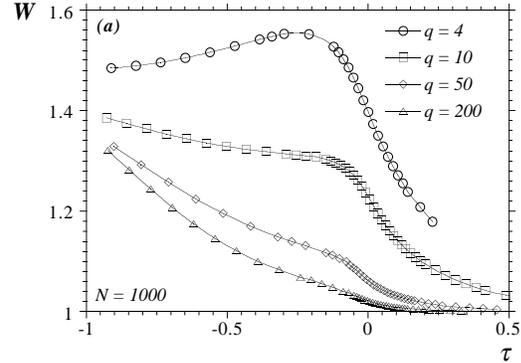

Figure 7. Plots of $W$ for various Potts models in the neighborhood of their critical points.

---
[2]In the simulations we have allowed degenerate triangulations with 2-loops and 1-loops.

It is interesting that the same picture has been suggested for models with $c > 1$ in [23]. It was suggested that there was a region $1 < c < c_0$ where the models had the same back-reaction on gravity as the $c = 1$ model.

### 2.4. Renormalization group approaches

Since finite size scaling seems to work well in two-dimensional gravity coupled to matter, it is natural to try to use another standard technique like the renormalization group. This technique might help us to define the scaling limit in situations where we have no analytical results to guide our choice of scaling. If one wants to apply the renormalization group equations there are

410

two basic steps:

(1) define large geometrical cells.

(2) define block spins or blocked matter fields on the larger geometrical cells.

Usually (1) is considered trivial and all the ingenious efforts enter into step (2). When we discuss dynamical triangulations (2) becomes trivial compared to (1). Somehow we have to find in a consistent way a method of forming blocks which preserves the underlying fractal structure of space-time. An old suggestion which goes all the way back to the study of hypercubic random surface models [24] is conceptually clear: Consider dynamical triangulations in two dimensions. The fundamental lattice length is the link length. It appears in the fractal structure precisely by the appearance of baby-universes where the neck-length is minimal, i.e. one if we allow one-loops in the triangulations. There will be a distribution of such baby universes as a function of their area. Compare this distribution to one where the class of triangulations does not allow one-loops, but does allow two-loops. Effectively this corresponds to a doubling of the cut-off length. In theory one can continue this blocking and the continuum fractal structure should be defined from the distribution of baby universes which represents a fixed point in the blocking procedure. From a practical point of view this procedure is useless. In [25] (see also contribution to this conference) it was noticed that the above idea could be drastically simplified if one stayed with a fixed neck-length of the baby universes, e.g. the minimal one. The distribution of baby universes of area $B$, for a fixed area $A$ of the total two-dimensional universe, is given by:

$$n(B) \sim AB^{\gamma_s - 2}. \tag{23}$$

However, the baby universes have a hierarchic structure in the case of spherical topology: One can cut them away, starting from the outer baby universes, and at each step close the boundaries. In this way the initial surface of area $A$ will lead to a sequence of ensembles of surfaces with areas $A < A_0$, the starting area:

$$\delta(A - A_0) \to P_1(A|A_0) \to P_2(A|A_0) \to \cdots \tag{24}$$

The distributions $P_1, P_2, \ldots$ are readily available from the simulations and this allows us to determine $\gamma_s$ under the hypothesis that the distribution of baby universes are given by (23).

A completely different approach has been suggested in [26,27]. The idea is to perform a blocking of a random lattice in such a way that some important geometric features are approximately preserved. Start out with a regular lattice and perform a blocking to a coarser lattice. Again we restrict ourself to 2d lattices for simplicity. The updating of the lattice structure is performed on the fine lattice. Mark the vertices of the coarser lattice. These vertices have an existence on the fine lattice. During the Monte Carlo simulation they will "float" around. How to update the coarse lattice, i.e. how to change neighbor assignment. In [26] it was suggested to let the link-distance on the underlying fine lattice determine whether or not to flip a link on the coarse lattice. This is illustrated on fig. 8: Two triangles $abc$ and $acd$ have the link $ac$ as common link. While the vertices $abcd$ have an existence on the finer lattice the links $ab$, $ac$ etc. only exist on the coarser lattice. The standard move on a two-dimensional triangulation is a link flip: $ac \to bd$. In order to decide if one should perform this flip the link distance between $a$ and $c$ and between $b$ and $d$ is calculated on the underlying fine lattice. If it is smaller between $b$ and $d$ the link is flipped from $ac$ to $bd$. The blocking of matter fields which are assigned to the vertices is straight forward.

In [27] it was shown that the prescription works well in pure gravity. If one adds formally irrelevant operators like $R^2$ to the Lagrangian and perform the blocking there is a clear flow to pure gravity.

It is important to make the above methods into genuine quantitative methods which allow us to extract critical exponents and correlation length in a reliable way.

2.5. Algorithmic improvements

In any Monte Carlo simulations the efficiency is intimately linked to the importance sampling. Whenever one encounter the phenomena of critical slowing down it is important to choose a sampling of configurations which is in accordance with the critical fluctuations. If the nature of the critical fluctuations is understood it might be possible to diminish the autocorrelation time between successive configurations by orders of magnitude. The best known example is the cluster algorithms for spin systems. These can in fact be used without any modifications on dynamical triangulations. For large central charge $c$ of the matter fields coupled to 2d quantum gravity the fractal structure becomes very pronounced as described above. The flip which is used to move around in the class of triangulations with fixed area becomes quite inefficient. This makes the computer simulation of large systems impossible. Recently it has been shown how the autocorrelation time for the dynamical triangulations can be drastically improved by using so-called *baby universe surgery* [28]. In the case of pure two-dimensional gravity one can cut away any baby universe outgrowth which is connected to the rest of the universe by a minimal bottle-neck and glue it back at any other position at the surface. This operation will not change the action at all, but if the baby universe is large it will be a global move which can only be imitated by many local moves. In addition it is perfect in accordance with the pronounced fractal structure of the surfaces for large $c$. In this way it has a lot in common with the cluster algorithm for spin systems. Using this algorithm it will be possible in the future to simulate systems of order $10^6$ triangles. This would be very time consuming relying entirely on local moves like the flip.

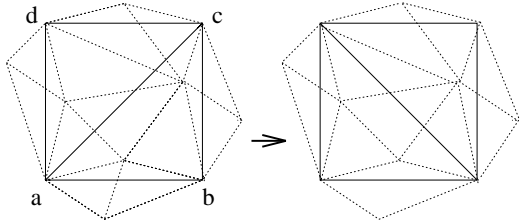

Figure 8. The flip of a link in the coarse grained lattice. The link distance between $a$ and $c$ in the underlying fine lattice is 3, while it is only 2 between $b$ and $d$.

There is clearly a lot of room for algorithmic improvement of the Monte Carlo simulations of dynamical triangulations and the baby universe surgery algorithm is presumably just the first of many improved algorithms.

## 3. QUANTUM GRAVITY FOR d>2

### 3.1. Definition of the model

Could it be that the formalism of simplicial gravity, which works so well for 2d gravity can serve as a definition for quantum gravity in $d = 4$? We are clearly moving into unchartered territory since there exists presently no continuum theory of quantum gravity in $d = 4$. The Euclidean path integral is not particularly well defined: First the Einstein-Hilbert action is unbounded from below: For any given metric assignment to a manifold, $g_{ab}(x)$, a conformal transformation

$$g_{ab}(x) \to \Omega(x)^2 g_{ab}(x) \qquad (25)$$

can make $R(x)$ arbitrary positive if only the derivatives of $\Omega(x)$ are sufficiently large. In addition we have to address the question of summation over topologies in the path integral

$$Z = \sum_{\mathcal{M} \in Top} \int_{\mathcal{M}} \frac{\mathcal{D}g_{ab}}{\text{Vol(diff)}} \, e^{-S_{eh}[g;\mathcal{M}]}. \qquad (26)$$

For $d \leq 3$ there is no discussion of the meaning of $Top$. The concept of homeomorphism is in one-to-one correspondence with the concept of diffeomorphism. However, for $d > 3$ this correspondence breaks down: there are topological structures which allow infinitely many smooth structures and there are topological structures which do not allow any smooth structures.

If we decide that the $\sum_{Top}$ means the summation over smooth structures it is possible to replace "smooth" by "piecewise linear" or "combinatorial inequivalent". For $d < 7$ there there is a one-to-one correspondence between "piecewise linear" and smooth structures. In this way simplicial gravity as described in the introduction offers a unique possibility to write eq. (26) in a

united way:

$$Z(k_2, k_4) = \sum_T e^{-k_4 N_4(T) + k_2 N_2(T)}. \qquad (27)$$

where the summation is over all combinatorial inequivalent abstract triangulations.

## 3.2. The entropy bound for fixed topology

Although the definition (27) is tantalizing from the point of view that it unites topology and metric structures in a natural way, it is unfortunately formal. This is true even in $d = 2$. The sum is divergent. A limiting procedure should be taken in order for sum to make sense. In two dimensions the so-called *double scaling limit* is an attempt to perform the summation.

In $d = 4$ it is not yet known how to perform the summation. However, it is generally believed that eq. (27) makes sense if one fixes the topology and the first numerical simulations seemed to support this idea [29]. A necessity is that the number of abstract triangulations of a given combinatorial manifold is exponentially bounded as a function of $N_4$, the number of four-simplexes. This is seen in the following way: If the topology is fixed it is easy to show that for any triangulation $T$ with $N_4(T)$ large we have

$$\frac{1}{2} N_4(T) \lesssim N_2(T) \lesssim \frac{10}{3} N_4(T). \qquad (28)$$

This implies that $Z(k_2, k_4)$ defined by (27) exists for a fixed topology if and only if there exists a $\bar{k}_4$ such that

$$\mathcal{N}_{\text{top}}(N_4) \equiv \sum_{T_{N_4} \in \text{top}} \leq e^{\bar{k}_4 N_4}. \qquad (29)$$

For a given $k_2$ the lowest possible $\bar{k}_4(k_2)$ is the critical value of the bare cosmological constant $k_4$.

Until now the only topology which has been used in the computer simulations is that of $S^4$ and the bound (29) has recently been questioned [30]. However, successive simulations clearly favor an exponentially bound on the number of triangulations [31,32]. Very recently an analytical proof of the exponential bound has been published [33].

## 3.3. Computational ergodicity

It is known that combinatorial four-manifolds are not algorithmically classifiable. This makes it questionable that one will ever be able to simulation the complete sum (27) numerically even if we manage to define it in one way or another. It is also known that there exists four-dimensional manifolds which are not algorithmically recognizable in the class of all four-dimensional combinatorial manifolds. Does this imply anything for Monte carlo simulations on such manifolds? This question was answered affirmatively two years ago by Ben-Av [34].

First a few definitions: Two triangulations are called combinatorial equivalent if there exists a triangulation which is a common subdivision of the two. The Monte Carlo simulations use a local update of the triangulations, which only requires a finite, fixed number of operations for each step. The updates are called *moves*. In addition the finite set of moves used are *ergodic*: Given two combinatorial equivalent triangulations $T_1$ and $T_2$ one can get from one to the other in a finite number of steps.

In case we consider triangulations of a manifold $\mathcal{M}_0$ which is not computational recognizable one can prove that if

$$N_4(T_1) < N \quad \text{and} \quad N_4(T_2) < N \qquad (30)$$

*the number of steps in the finite algorithm to get from any triangulation $T_1$ of $\mathcal{M}_0$ to any other triangulation $T_2$ of $\mathcal{M}_0$ cannot be bounded by a recursive definable function $r(N)$.* This includes functions like $N!$ and $N!^{N!}$. For a given $N$ the total number of triangulations with $N_4(T) < N$ is clearly bounded by some $(aN)!$ where $a$ is of order one. How can the above mentioned theorem be true? It can only be true if we on the way from some $T_1$ to some other $T_2$ is forced to triangulations with very large $N_4$. In this way there will be barriers separating different regions of configuration space. In fact one can prove that the height of the barrier cannot be bounded either by a recursively definable function.

Although in principle ergodic, the class of local moves mentioned above cannot in practise reach all triangulations of $\mathcal{M}_0$. However, it is not known if the region of configuration space we can-not reach is of measure zero in the limit $N \to \infty$.

How is the situation for $S^4$, which is the topology used until now. It is not known if $S^4$ is algorithmically recognizable or not. Extensive computer simulations looking for any sign of barriers have failed (see [35] or the contribution to Lattice 94). This can be taken as a hint that either $S^4$ is computational recognizable or a hint that the regions which cannot be reached are of measure zero. It would be most interesting to perform the simulations for a manifold which is known not to be algorithmically recognizable.

### 3.4. Scaling

The following scenario seems to be true in four-dimensional simplicial gravity: For large values of the bare gravitational coupling constant (small values of $k_2$) the typical universes are very crumpled with almost no extension if distance is measured as the shortest link length between two vertices. For small values of the gravitational coupling constant (large values of $k_2$) the universes seem very elongated, and the Hausdorff dimension seems close to two: the universes seem to be branched polymers!. There is a phase transition, most likely of second or higher order where the Hausdorff dimension might be close to four!.

The different groups which have performed computer simulations agree more or less on this scenario. The next step is to analyze in detail the scaling behavior in the infinite volume limit close to the phase transition. The first step has been taken in [36] where it was found that distribution of internal geodesic distances seems to allow a scaling which indeed indicated a flow towards the transition point in the following way: The distributions of distances for universes corresponding coupling constant away from the critical point could be mapped on distributions for large universes close to the transition point. This is the first genuine hint of a divergent correlation length when we approach the critical point.

Clearly much remain to be done. In particularly it would be helpful with simulations of larger universes. Until now most simulations have used $N_4 < 32.000$. Although a large number such $N_4$'s correspond in fact to quite small four-dimensional manifolds.

### 4. Summary

Two-dimensional gravity is a wonderful playground for both numerical and analytical work. Many of the tools we have available from statistical mechanics seems to work well and sometimes the models formulated on dynamical triangulations are much more easily solved analytically than the counter parts defined on a regular lattice. This might be a reflection of the underlying reparametrization invariance of the theory in the scaling limit.

Higher dimensional gravity is still unchartered territory. Hopefully the numerical simulations will teach us something about the typical quantum universe and inspire analytic work. Dynamical triangulations have proven a very powerful tool in two dimensions and one can hope that the same will be true in higher dimensional gravity.

There is a lot of room for algorithmic improvements in the simulations. The first step has been taken by the use of baby universe surgery. It would be very helpful if we could increase the size of the simulated universes by some order of magnitudes.